\documentclass[a4paper]{jpconf}
\usepackage{graphicx}
\usepackage{wrapfig}
\usepackage{subfigure}

\begin{document}

\title{Directed Flow of Charged Kaons in Au+Au Collisions from the BES Program at  RHIC}

\author{ Yadav Pandit (for the STAR Collaboration)}

\address{Department of Physics, UIC, Chicago, USA}

\ead{ypandit@uic.edu}

\begin{abstract}

 We report the measurement of  the directed flow ($v_{1}$) for charged kaons in  Au+Au collisions at  $\sqrt{s_{NN}}$ =7.7, 11.5, 19.6, 27, 39, 62.4 and 200 GeV as a function of rapidity and compare these results for pions, protons and antiprotons.  These new kaon results may help to constrain the medium properties and collision dynamics including the in-medium kaon potential and baryon number transport in these collisions.
\end{abstract}

\section{Introduction}

 The study of collective flow in relativistic nuclear collisions has potential to offer insight into the equation of state of the produced matter. Directed flow, $v_1$, is the first harmonic coefficient of the Fourier expansion of the final-state momentum-space azimuthal distribution \cite{methods}, 
\begin{equation}
v_1 = \langle \cos ( \phi-\Psi_R ) \rangle, 
\end{equation}
 \begin{figure}[h]
 \center
\includegraphics[width=35pc]{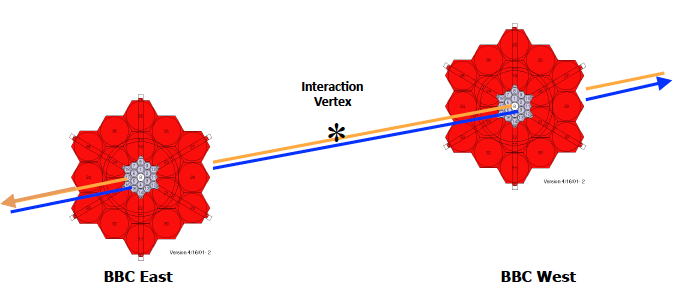}
\caption{ Beam Beam Counters at the STAR detector}
\label{fig1}
\end{figure}
where the angle brackets indicate an average over all the particles and events used, $\phi$ denotes the azimuthal angle of the outgoing particles and $\Psi_R$ is the orientation of the event plane which is determined event-by-event. Both hydrodynamic~\cite{Hydro} and nuclear transport models~\cite{urqmd} indicate that directed flow is a promising observable for investigating a possible phase transition, especially in the region of Beam Energy Scan (BES) program at RHIC.  In particular, the shape of $v_1$ as a function of rapidity, $y$, in the mid-rapidity region is of interest because it has been argued that it is sensitive to crucial details of the expansion of the participant matter during the early stages of the collision.  The models indicate that the evolving shape and orientation of the participant zone and its surface play a role in determining the azimuthal anisotropy measured among these particles in the final state~\cite{prequilibrium,Heinz}. 

 \begin{figure}[h]
 \center
\includegraphics[width=35pc]{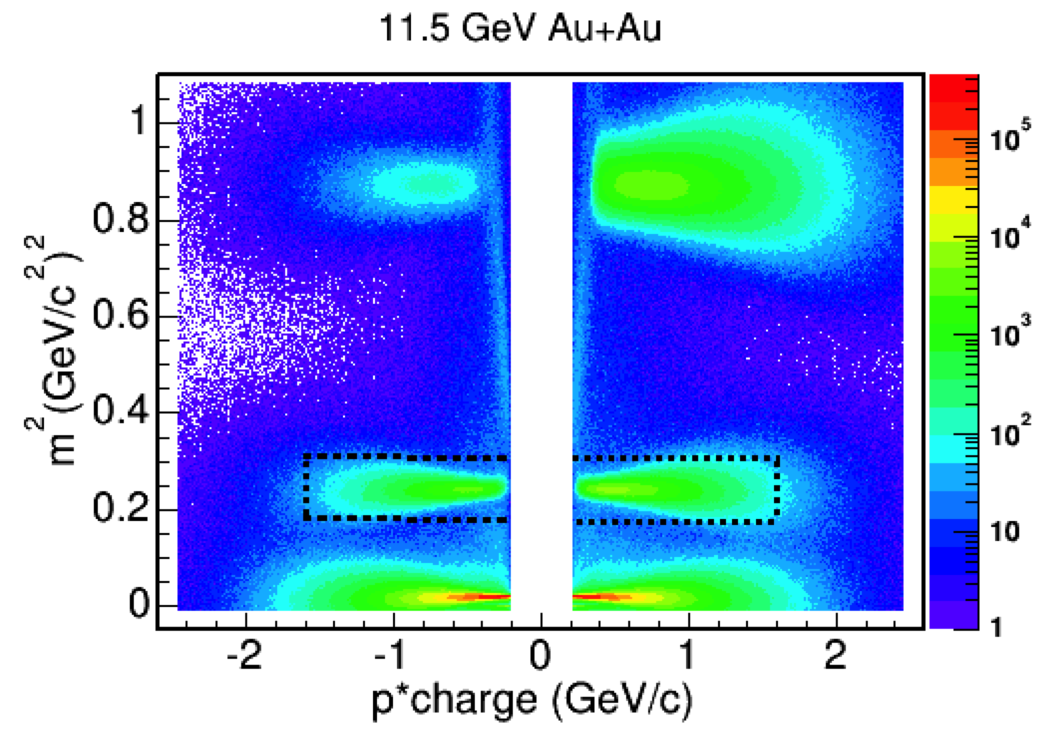}
\caption{ Mass squared based on time-of-flight information, versus momentum from TPC curvature, for the case of Au+Au collisions at 11.5 GeV. $K^{+}$ and $K^{-}$ candidates are indicated by dashed-line boxes.}
\label{fig2}
\end{figure}

It is generally believed that the directed flow is generated early in the heavy-ion collision. Therefore it may be an important probe to provide valuable information of  the onset of bulk collective dynamics during thermalization, even on the pre-equilibrium stage. Recently STAR published  the  measurement of directed flow of identified hadrons (protons, antiprotons, and
positive and negative pions) with high precision for semi-central Au+Au collisions in the energy range $\sqrt{s_{NN}}  =  7.7-200$ GeV~\cite{PRLSTAR}. After the publication of these data, there have been some theoretical efforts \cite{FrankFort,PHSD} to understand the underlying physics and this is an ongoing effort. Still sizable discrepancies with experimental measurements in the directed flow characteristics are found for the microscopic kinetic models at $\sqrt{s_{NN}} < $ 20 GeV  and are common for the transport models HSD,  PHSD and UrQMD.  There are few other observables where kinetic approaches have a problem in this energy range ~\cite{PHSD}.
 \begin{figure}[h]
 \center
\includegraphics[width=25pc]{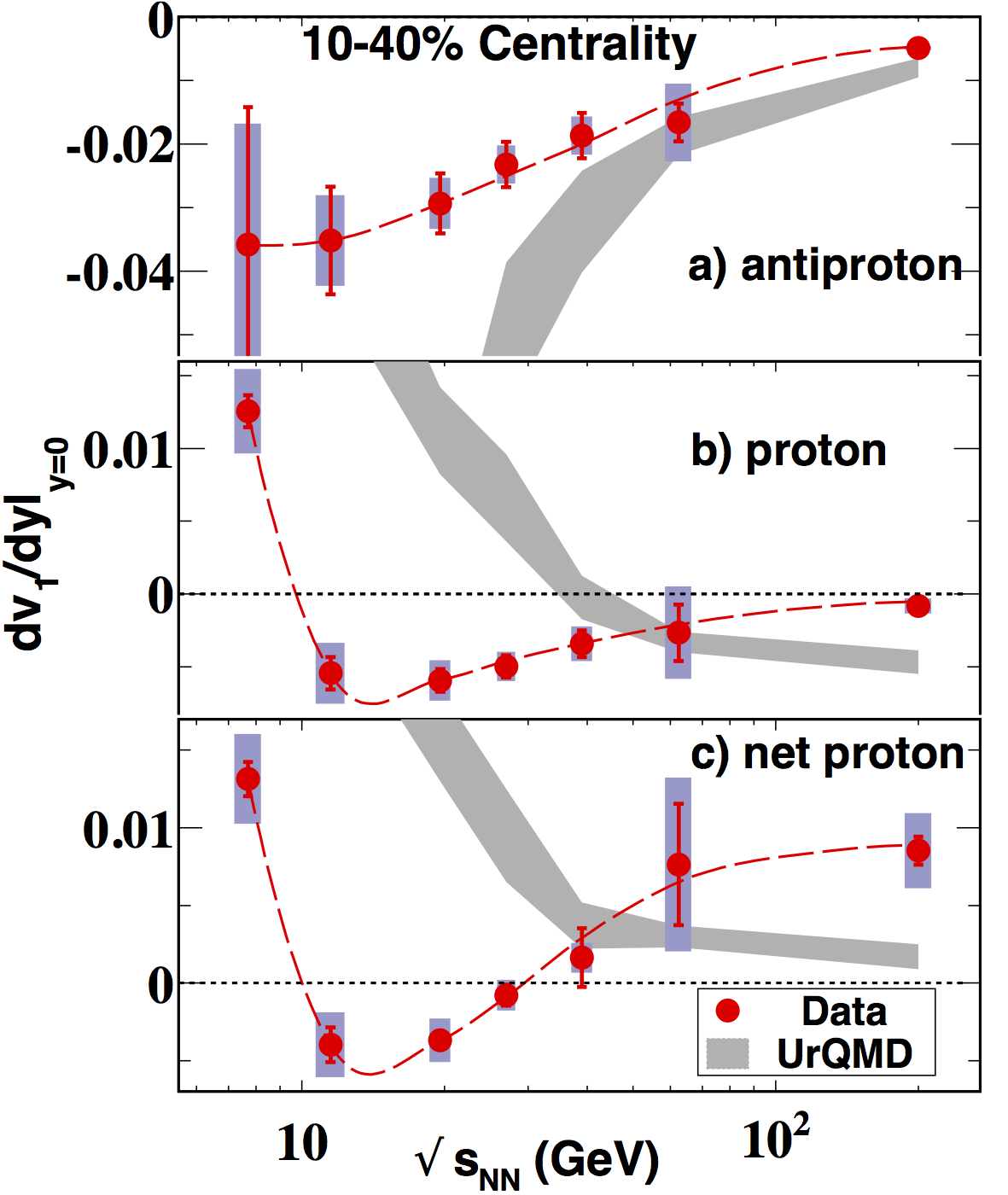}
\caption{ Directed flow slope ($dv_1/dy$) near mid-rapidity versus beam energy for intermediate-centrality Au+Au collisions.  The slopes for protons, antiprotons and net-protons  are reported ~\cite{PRLSTAR}.}
\label{fig3}
\end{figure}

The role of transported baryon number and absorption effects are not well understood. There are still a lot to be done, specially in the model calculation since the effect of event plane used in the experiment does not exactly coincide with the known reaction plane in the theory calculation. The role of produced particle vs spectator in the event plane calculation may change with energy since the event plane detector is in fixed phase space. UrQMD simulation suggests that these effects will have a minimal effect in the experimental results. 
   
    We report here the directed  flow  $v_{1}$  measurements for charged kaons in Au+Au collisions from the STAR  experiment at the Relativistic Heavy Ion Collider (RHIC)  in the range $\sqrt{s_{NN}}=$ 7.7 to 200 GeV, based on data from STAR experiment collected during RHIC run  in the years 2010 and 2011. These measurements are presented as a function of rapidity and centrality.  

 \section{The STAR experiment and Analysis Details}
 
 \begin{figure}[h]
 \center
\includegraphics[width=26pc]{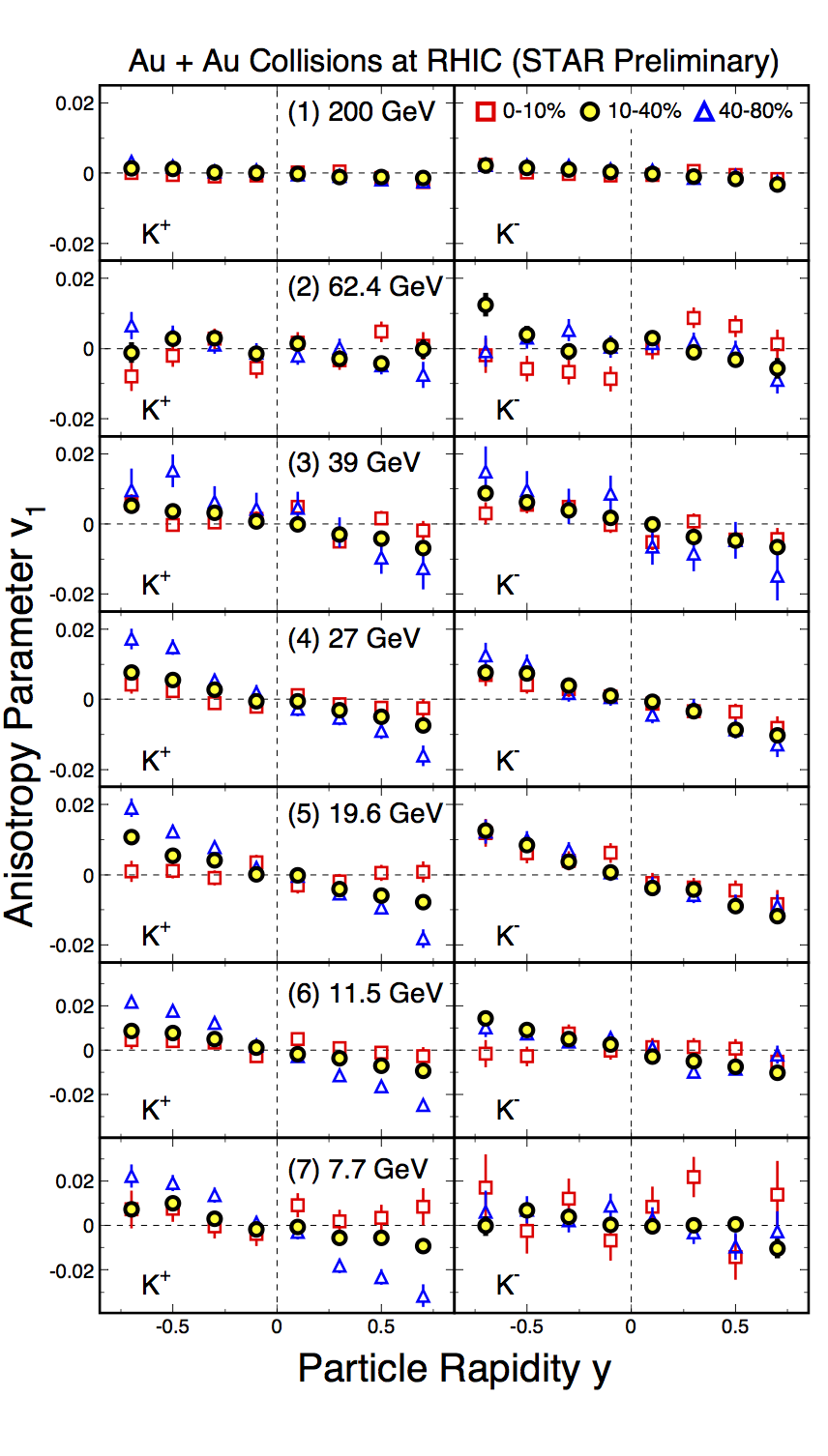}
\caption{ Directed flow for charged kaons  $K^{+}$ on left panel  and $K^{-}$ on right panel  versus rapidity for central (0-10\%), intermediate-centrality (10-40\%) and peripheral (40-80\%) Au+Au collisions at $\sqrt{s_{NN}}$= 7.7-200 GeV.  The plotted errors are statistical only, and systematic uncertainties are explained in the text.}
\label{fig4}
\end{figure}

    
 Data reported in these proceedings were collected in Au+Au collisions at $\sqrt{s_{NN}}$ = 7.7- 200 GeV in the year 2010 and 2011 with a minimum bias trigger.  The main detector used in the present measurements is the Time Projection Chamber (TPC)~\cite{startpc},  the primary tracking device at STAR, time of flight detector and beam beam counter . TPC has  full azimuthal coverage and uniform acceptance in $\pm 1.0 $ units of pseudorapidity.   The charged particle momenta are measured by reconstructing their trajectories  through the TPC.  The centrality was determined by the number of tracks from the pseudorapidity region $ |\eta|  < 0.5$.
 The Beam Beam Counter (BBC) detector subsystem ~\cite{BBC}  consists of two detectors mounted around the beam pipe, each located outside the STAR magnet pole-tip at opposite ends of the detector approximately 375 cm from the center of the nominal IR as shown in Fig 1.   Each BBC detector consists of nearly circular scintillator tiles arranged in four concentric rings that provide full azimuthal coverage.  The two inner rings have a pseudorapidity coverage of $3.3 < |\eta| < 5.0$ and are used to reconstruct the first-order event plane for this analysis.  Charged kaons  were identified  up to 1.6 GeV/$c$  in transverse momentum  based on specific energy loss in the TPC and from the time measurement by STAR's time-of-flight barrel in combination with the momentum. 
      
  Possible systematic uncertainties arising from non-flow, i.e., azimuthal correlations not related to the reaction plane orientation (arising from resonances, jets, strings, quantum statistics, and final-state interactions like Coulomb effects) are reduced due to the relatively large pseudorapidity gap between the STAR TPC and the BBC detectors~\cite{methods}. Directed flow measurements based on the BBC event plane, where the BBC east and west detectors ensure symmetry in rapidity acceptance, cancel biases from conservation of momentum in the basic correlation measurement because the difference in $v_1$ between the rapidity hemispheres is used. However, momentum conservation effects  do contribute to systematic uncertainty in the event-plane resolution, and thereby in the resolution-corrected signal, at the level of less than 2\% ~\cite{PRLSTAR}.   The systematic uncertainty arising from particle misidentification and detector inefficiency is estimated by varying event and track cuts, and is $\sim 5$\%.

\section{Result and discussion}
 \begin{figure}[h]
 \center
\includegraphics[width=35pc]{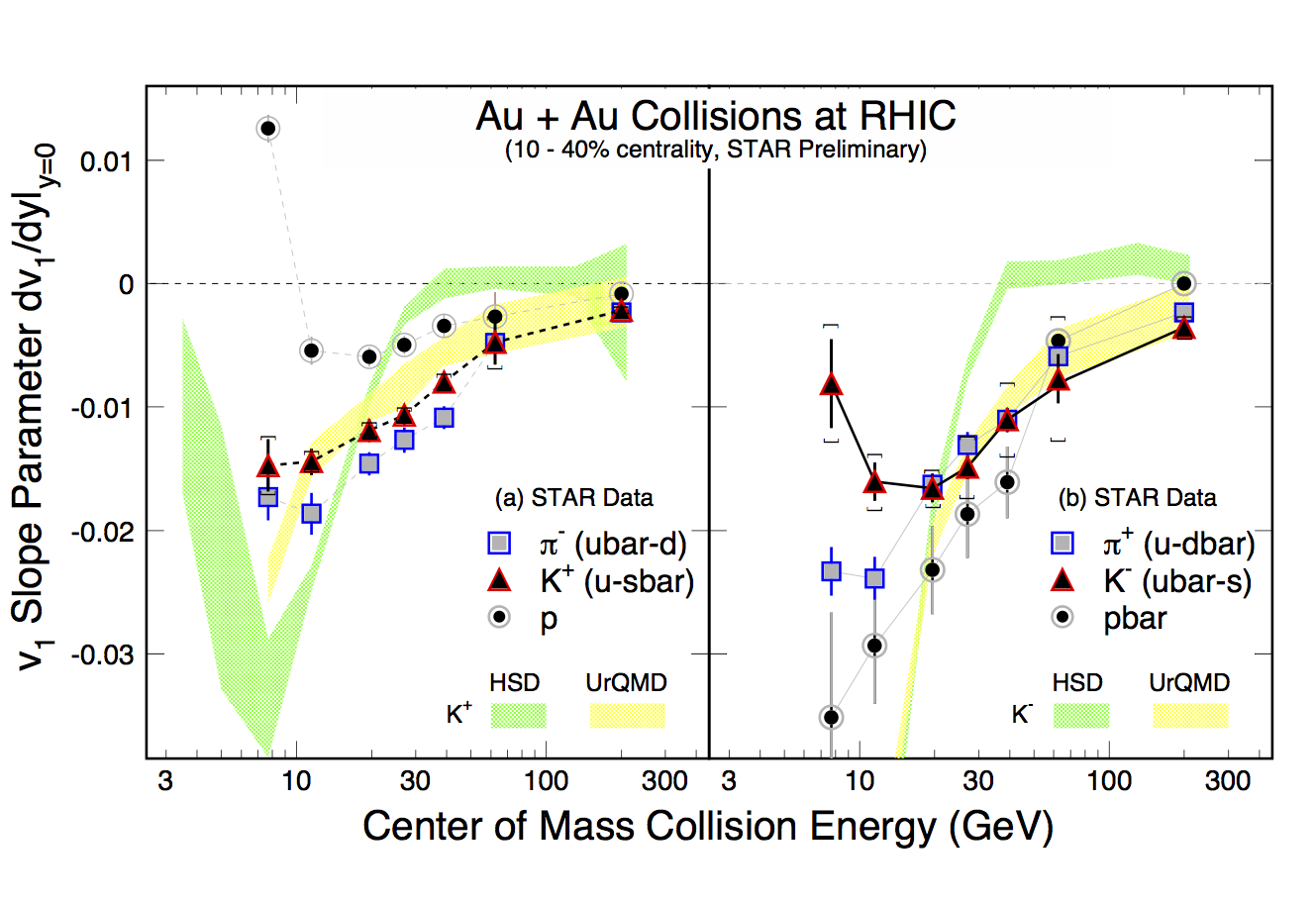}
\caption{ Beam energy dependence of the slope $dv_{1}/dy$ at mid-rapidity for identified hadrons for intermediate centrality (10-40\%). New kaon measurements are shown as filled triangles, while results for protons, antiprotons and pions were previously reported in Ref.~\cite{PRLSTAR}. All slopes are based on a linear fit. Statistical and systematic errors are represented by bars and caps, respectively. Calculations from Hadron String Dynamics (HSD)~\cite{PHSD}  and UrQMD~\cite{FrankFort} are shown as shaded bands.}
\label{fig5}
\end{figure}

In Fig.~\ref{fig4}, directed flow for charged kaons,  $K^{+}$ on left panel  and $K^{-}$ on right panel  versus rapidity for central (0-10\%), intermediate-centrality (10-40\%) and peripheral (40-80\%) Au+Au collisions at $\sqrt{s_{NN}}$= 7.7-200 GeV are shown.  All data points in Fig.~\ref{fig4} are antisymmetric about mid-rapidity, verifying cancellation of the momentum conservation effect\cite{MomentumCons}. In intermediate and peripheral collisions, slopes of $v_1(y)$ near mid-rapidity for $K^{\pm}$  are generally negative for all energies with larger errors.  Furthermore, STAR has previously reported negative slopes at mid-rapidity for pions for the energy range reported here and for protons the slopes were negative for 11.5 and higher beam energies ~\cite{PRLSTAR}.

Fig.~\ref{fig5} reveals the  beam energy dependence of the directed flow slope $dv_{1}/dy$ for $K^{\pm}$ near midrapidity for intermediate centrality (10-40\%) along with protons, antiprotons and $\pi^\pm$.  On the left panel we show results for $K^{+}$ compared with $\pi^{-}$ and protons. On the right panel we compare $K^{-}$ with $\pi^{+}$ and antiproton.  New kaon measurements are shown as filled triangles, while results for protons, antiprotons and pions were previously reported in Ref. ~\cite{PRLSTAR}. All slopes are based on a linear fit. Statistical and systematic errors are represented by bars and caps, respectively. Calculations from Hadron String Dynamics (HSD) ~\cite{PHSD}  and UrQMD ~\cite{FrankFort} are shown as shaded bands. At the lowest collision energy, $\sqrt{s_{NN}}=$ 7.7 GeV, charged kaons and all other produced hadrons show opposite sign of the mid-rapidity slope parameter compared to protons. In this low energy region, $K^{ - }$  $v_{1}$ slope is closer to zero  than that of $K^{ + }$.  This observation supports the expectation that  $K^{+}$  experiences a repulsive and $K^{-}$ experiences attractive nuclear  potential at these energies where baryon density is high. An extrapolation of STAR's systematic energy dependence is consistent  with fixed-target results from the AGS.  At  the higher energy region, $\sqrt{s_{NN}} > $ 30 GeV, where  pair production becomes dominant, all observed particles seem to show a similar beam energy dependence, and the difference among them reduces as the energy increases. 
          
\section {Summary}

In these proceedings, STAR results for directed flow $v_{1}(y)$ and directed flow slope ($dv_{1}/dy$) for charged kaons as a function of center of mass colliding energy and collision centrality are presented. At the lowest collision energy, $\sqrt{s_{NN}}=$ 7.7 GeV, charged kaons and all other produced hadrons show opposite sign of the mid-rapidity slope parameter compared with protons. At  the higher energy region, $\sqrt{s_{NN}} > $ 30 GeV, all observed particles seem to show a similar beam energy dependence, and the difference among them reduces as the energy increases. The transport model HSD, with the mean-field on, and UrQMD provide a reasonably good comparison to data at the high energy region, but fail at low energy. The present measurements might be helpful to further constrain the medium properties in terms of mean-field and the interplay of quark and baryon transport.

\end{document}